\documentstyle[11pt,newpasp,twoside,natbib,graphics]{article}
\def\edcomment#1{\iffalse\marginpar{\raggedright\sl#1\/}\else\relax\fi}
\reversemarginpar

\begin{document}
\setcounter{page}{265}
\title{Radiative Transfer Modeling of Passive Circumstellar Disks: Application to HR 4796A}
  \author{Thayne Currie}
\affil{Department of Physics and Astronomy, University of California-Los Angeles, Los Angeles, CA 90095}
\author{Dmitry Semenov}
\affil{Astrophysical Institute and University Observatory, Schillergafchen 2-3, 07745 Jena, Germany }
\author{Thomas Henning}
\affil{Max Planck Institute for Astronomy, Konigstuhl 17, 69117 Heidelberg, Germany}
\author{Elise Furlan and Terry Herter}
\affil{Department of Astronomy, Cornell University, Ithaca, NY 14853}
\begin{abstract}
 We present a radiative transfer model which computes the spectral
 energy distribution of a passive, irradiated circumstellar disk,
 assuming the grains are in radiative equilibrium.  Dependence on the
 radial density profile, grain temperature estimation, and optical
 depth profiles on the output SED are discussed.  The best fit model
 for HR 4796A has a minimum and maximum spherical grain size of 2.2
 and $1000\ \mu$m respectively, a size distribution slightly steeper
 than the "classical" -3.5 MRN power law, grains composed of
 silicates, trolite, ice, and organics and a peak radial density of
 $1.0\times10^{-17}$ g ${\rm cm}^{-3}$ at 70 AU, yielding a disk mass
 of roughly $2M_{\oplus}$.
\end{abstract}

\section{Introduction}
Modeling of the spectral energy distribution (SED's) of circumstellar
disks can potentially yield a wealth of information about the system
under investigation.  Sophisticated disk models have recently been
developed by Chiang and Goldreich (1997) and Wolf (1999) for T Tauri
disks, and basic modeling of older, optically-thin Vega-like disks was
performed by Artymowicz (1989) for $\beta$ Pictoris and Augereau et
al. (1999) (hereafter, AG 99) for HR 4796A.

We present a radiative transfer model of passive, Vega-like dusty
disks that more rigorously treats the optical properties of disk
grains and emergent intensity from along the observer's line of sight
than AG 99, allowing for better constraints of grain composition,
surface density profiles, and disk mass.  HR 4796A was chosen as a
test system because of the large number of infrared excess flux
measurements in the near infrared to submillimeter regions of the
spectrum (summarized in Table 1) and restrictions on disk geometry,
providing tighter constraints for disk model input parameters.

\begin{table}
\caption{Infrared and submillimeter flux measurements for HR 4796A used to constrain the model.}
\vspace{2mm}
\begin{tabular}{lrll}
\hline
{$\lambda$} & {Excess Flux (Jy)} & {Uncertainty (Jy)} &{Source} \\
\hline
11.6 & 0.086  &   0.070 & Fajardo-A. et al. (1998) \\
12.5 & 0.101  &   0.018 & Koerner et al. (1998) \\
12.5 & 0.133  &   0.027 &Fajardo-A. et al. (1998)  \\
18.2 & 1.100  &   0.150 &Jayawardhana et al. (1998)  \\
20.0 & 1.860  &   unknown &Jura et al. (1993)  \\
20.8 & 1.813  &   0.170 &Koerner et al. (1998)  \\
24.5 & 2.237  &   0.700 &Koerner et al. (1998)  \\
25.0 & 3.250  &   0.130      &IRAS  \\
60.0 & 8.630  &   0.430      &IRAS  \\
100.0 &4.300  &   0.340      &IRAS  \\
450.0 &0.180  &   0.150 &Greaves et al. (2000)  \\
800.0 &$<0.028$&         &Jura et al. (1995)  \\
850.0 &0.0191&    0.0034 &Greaves et al. (2000)  \\
\hline
\end{tabular}
\vspace{-2mm}
\end{table}

\section{Passive Circumstellar Disk Modeling Theory}
The emergent flux from a passive, optically thin circumstellar disk at
some wavelength equals the wavelength dependent intensity of light
emerging from a point in the disk integrated over the solid angle that
the disk comprises:
\begin{equation}
F_{\nu}= D^{-2}\int_0^{2\pi}\int_0^{r}I_{\nu}r_{m}sin(i)  dr_{m}d\omega
\end{equation}
where D is the star's distance from the Earth, $I_{\nu}$ is the
emergent intensity at a point in the disk with midplane radius
$r_{m}$, and $i$ is the inclination angle where $i = 90$ corresponds
to a face-on disk.  The intensity emergent from each point in the disk
is calculated by solving the radiative transfer equation and
integrating the intensity contribution from all grains along the line
of sight:
\begin{equation} 
I_{\nu}(r)= \int_0^{s} B_{\nu}(T)\kappa(\lambda)\rho(r)ds
\end{equation}
where $\kappa$ is opacity, and $\rho$ is mass density of grains.  The
grains' effective temperatures, $T_{g}$, are computed assuming dust
grains are in radiative equilibrium:
\begin {equation}
q_{uv}(R_{*}/r)^{2}\sigma T_{*}^{4} = 4\int_0^{\infty} q_{ir} B_{\nu}(T_{g})d\nu
\end {equation}
where $q_{uv}$ (early A stars like HR 4796A emit mostly UV photons)
and $q_{ir}$ are respectively the absorption efficiencies for incident
and emitted radiation.  Because grain extinction coefficients are
wavelength dependent, and, in more rigorous treatments, cannot be
expressed as a simple analytic function, one must numerically solve
for the grain temperature.  Grain temperatures are calculated over a
grid of grain sizes and distances.  Resolved images of the HR 4796A
disk infer a radial density profile sharply peaked at 70 AU of the
form $\rho(r)=\rho_{o}(r/rc)^{\gamma}$ where $rc$ = 70AU, $\gamma$ is
$\gamma_{in}$ for $r\leq rc$ and $\gamma_{out}$ for $r> rc$ are the
inner and outer radial density power law factors, $\rho_{o}$ is the
peak radial density, and $rm$ is the distance at which this occurs.
Wavelength-dependent opacities are calculated from the Semenov et
al. (2002) compositional model for chemically homogeneous, spherical
grains whose primary components are pyroxene, olivine, volatile and
refractory organics, water ice, trolite, and iron: similar to the
Pollack (1994) compositional model.  The optical properties of dust
particles are computed via Mie theory.  The grain size distribution is
initially set to the size distribution for interstellar grains as
prescribed by Mathis (1977).  Minimum size is set at 5 $\mu$m for HR
4796A: close to the blowout size limit as inferred by AG 99.  These
two parameters are then adjusted to yield an optimum SED fit.
\begin{figure}
{\centering \resizebox*{0.7\columnwidth}{!}{\includegraphics{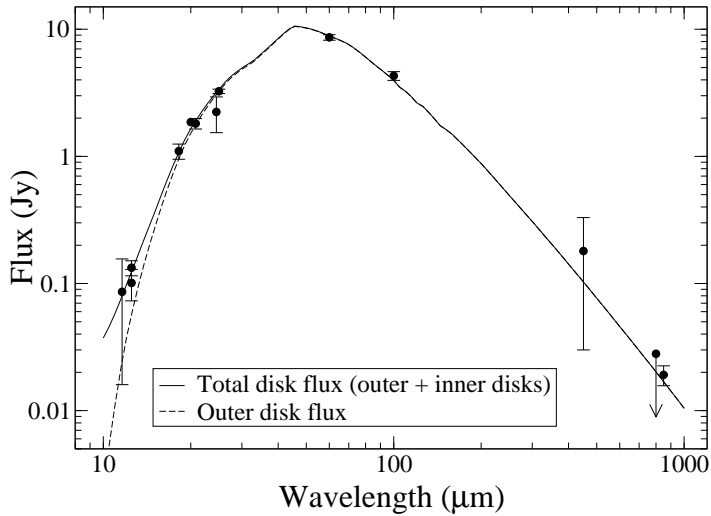}}\par}
\caption{Full synthetic SED for outer disk and total disk contribution compared
 against observed infrared excess measurements.}
\end{figure}
\section{Modeling Results}
Results of the disk modeling effort are shown in Figure 1.  Optimum
fit of the SED was achieved with the following input parameters:
$a_{min}$ = 2.2 $\mu$m, $\beta$ = -3.86, $\gamma_{in}$ = 10,
$\gamma_{out}$ = -12, and $\rho_{o} = 1.0\times10^{-17}$ g ${\rm
cm}^{-3}$.  An additional hot dust component peaked at 9 AU was
required to fit the SED.  Predicted flux values fall within the margin
for error of measured values at nearly all wavelength points.  The
best fit requires a minimum size smaller than the blowout size limit
for the system.  Thus, some of the dust grains are not primordial and
may be replenished via collisions of larger bodies.  This model gives
a disk mass of roughly $2.05 M\oplus$: consistent with dynamical
calculations of the disk mass (AG 99) and Greaves (2000) calculation
of $M_{disk} \geq 0.25 M_{\oplus}$ and does so without using scaling
factors or 'fitting parameters'.  The only true free parameter is the
peak radial density.  The radial density power laws can be
qualitatively inferred from resolved images of the disk.  The emergent
SED varies so wildly with the grain composition assumptions that only
a very small subset of possible compositions (which turn out to be
quite similar) will yield empirically accurate model predictions.
Finally, the gas/dust ratio, while not known, can be crudely
constrained, and from this type of modeling one can produce a grid of
models having a range of peak densities and gas/dust ratios.  From
Greaves (1999), the gas mass must be $< 7 M\oplus$ and the age and
optical depth of the HR 4796A disk suggest a small gas/dust ratio:
this is consistent with our results.

\section{Future Tasks}
The main remaining task is to make the absorption/extinction
coefficient treatment self-consistent.  While the grain optical
properties of the were treated rigorously in the opacity calculations,
grain temperatures themselves were calculated using a crude analytical
function characterizing the extinction coefficients for the grains.
While the derived grain temperatures concur roughly with previous
temperatures derived by AG 99, a self-consistent approach would ensure
more accurate grain temperatures.  Second, a mechanism for confining
HR 4796A's disk material into its narrow, sharply peaked annulus
should be sought.  Wyatt et al. (1999) suggested that orbital
resonances with a jovian mass planet interior to the main disk could
be responsible: reminiscent of the Goldreich and Tremaine (1982)
discussion of ring confinement by shepherding moons.  Finally, similar
models could be produced for Vega, a slightly older, early A-type star
which seems to be tracking an evolutionary path similar to that of HR
4796A, with results compared to previous modeling attempts for this
system.

\end{document}